\documentclass[aps,pre,preprint,showpacs,bibnotes]{revtex4}

\newcommand{\be}{\begin{equation}}
\newcommand{\ee}{\end{equation}}
\newcommand{\bea}{\begin{eqnarray}}
\newcommand{\eea}{\end{eqnarray}}
\renewcommand{\d}{\text{d}}

\usepackage{amsmath,amsfonts,bbm,mathrsfs,graphics}

\begin{document}

\title{Distance characteristics of ergodic trajectories in two-dimensional non-integrable systems: Reduced phase space}

\author{Jamal Sakhr}
\affiliation{Department of Physics and Astronomy, University of Western Ontario, London, Ontario N6A 3K7 Canada}

\date{\today}

\begin{abstract}
A fundamental process for any given chaotic flow is the deterministic point process (DPP) generated by any chaotic trajectory of the flow repeatedly crossing a canonical surface-of-section (herein referred to as a $\Sigma$-type DPP). This paper introduces the idea of using stochastic point process models to describe and understand the spatial statistical features of $\Sigma$-type DPPs in two-degree-of-freedom (2D) Hamiltonian systems. In the specific context of 2D non-integrable systems possessing ergodic components, it is proposed that, in an ergodic region, the pertinent model for describing the spatial statistical features of any typical $\Sigma$-type DPP is the two-dimensional homogeneous Poisson point process (herein denoted by $\mathbf{P}_2$). Of particular interest in this paper are the (Euclidean) $k$th-nearest-neighbor distance characteristics ($k=1,2,3,\ldots$) of a given $\Sigma$-type DPP. Employing the two-dimensional cardioid and semi-circular mushroom billiards as generic test cases, it is shown that typical sample $\Sigma$-type DPPs possess $k$th-nearest-neighbor distance characteristics consistent with model predictions for $\mathbf{P}_2$. Deviations from this observed Poissonian behavior are expected in strictly non-ergodic regions (i.e., the chaotic regions in generic Hamiltonian systems), but in cases where the dynamics is strongly (but not fully) chaotic, such deviations are posited to be negligible. The validity of the latter claim is demonstrated in the case of the 2D H\'{e}non-Heiles system at the critical energy. The results of the numerical experiments are contextualized and their significances to both classical and quantum chaos are discussed.  
\end{abstract}


\maketitle

\section{Introduction} 

Chaotic particle trajectories of classical conservative systems have an intricate and complex spatial structure. A key theoretical issue in the study of Hamiltonian systems (and in nonlinear dynamics more generally) is the characterization of this spatial complexity. Ideas and techniques from a range of mathematical subjects, including topology \cite{MarsdenMech,Pettini}, differential geometry \cite{Arnold,Vilasi,MarsdenGeo}, and fractal geometry \cite{Umberger85,benettin86,Hanson87,Jefferys87,Barrio2008}, have been profoundly useful in both identifying and describing many fundamental spatial features associated with classical particle trajectories. Computer simulations have also been instrumental in revealing the qualitative spatial structure of both regular and chaotic trajectories in a variety of paradigmatic model systems (see, for example, Ref.~\cite{SICM}). The spatial complexity frequently observed in computer simulations can, in part, be described using the language of fractal geometry, which has so far been the predominate mathematical paradigm for investigating questions concerning the geometrical structure of chaotic trajectories. There are however questions concerning the geometrical structure of chaotic trajectories that fall outside the scope of both differential and fractal geometry. One such question, of a statistical nature, is introduced below.   

The most prominent spatial attribute of the chaotic trajectories is the extreme spatial asymmetry that results from their seemingly unsystematic wandering through phase space. The seemingly haphazard evolution of these trajectories naturally elicits the use of stochastic modeling for the purpose of characterizing their spatial structure. Indeed, the conventional wisdom is that any evolving chaotic trajectory mimics some sort of random process (see, for example, Refs.~\cite{Zas93,Zas00,Zassy2002}). This immediately raises a fundamental question: What type of stochastic process(es) can aptly model the spatial characteristics of the chaotic trajectories in (for instance) two-degree-of-freedom (2D) Hamiltonian systems? Questions of this nature commonly confront practitioners of ``spatial analysis''. The idea of using stochastic processes to model the spatial structure of a spatially complex object (such as a chaotic particle trajectory) is actually one that is central to the mathematical fields of spatial and geometrical statistics \cite{Rips,StoyandStoy}. The question being raised is thus (in essence) one that falls within the purview of these disciplines. The words ``spatial characteristics'' (a verbalism frequently employed in these fields) will likely be ambiguous to some readers. To be more explicit, it is useful to recast the above-posited question in more direct terms as the following two-part question: What is the \emph{spatial statistical} (or \emph{geometrical-statistical}) structure of the chaotic trajectories of a 2D (non-integrable) Hamiltonian system, and what stochastic geometric model(s) \cite{SKM} (if any) can serve to elucidate this structure? To the author's knowledge, these questions have not been addressed in the literature nor have (more generally speaking) the ideas of spatial and geometrical statistics been explicitly applied to the study of Hamiltonian systems. 

As mathematical objects classical particle trajectories can be subjected to a wide range of spatial statistical analyses and so the general study of their overall spatial statistical structure is a multifaceted problem. The intent here is to introduce what is perhaps the most elemental sub-problem: determining the spatial statistical properties of the points on a trajectory that intersect any representative surface-of-section. Rather than immediately proceeding to the specifics of the problem, it is conceptually useful to formulate the basic questions involved using the language of point processes. In the nonlinear dynamics literature, a (deterministic) process which reduces the dynamics to a point set using a series of event timings is often called a ``point process'' (see, for example, Ref.~\cite{Sauer94}). Unfortunately, the term ``point process'' more commonly refers to a certain type of \emph{stochastic} process \cite{SKM}. Thus, following the authors of Ref.~\cite{DingTang}, the term ``\emph{deterministic} point process'' (DPP) will here be used in reference to any deterministic process of the type alluded to earlier, whereas the term ``point process'' shall here retain its traditional (stochastic) meaning. The successive intersections of any (continuous) phase space trajectory with a surface-of-section are successive point events that define a DPP. What are the spatial statistical properties of such a DPP? Is there a (stochastic) point process that can aptly model these properties? The spatial statistical properties of this particular type of DPP (henceforth referred to as a $\Sigma$-type DPP) will of course vary depending on the nature of the generating trajectories. (In mixed systems, for example, island chains will have different spatial statistical properties than trajectories which visit the chaotic sea.) In the present paper, the preceding questions shall be addressed in the specific context of 2D non-integrable Hamiltonian systems that are either fully ergodic or possess any number of ergodic components.  

The contents of the paper, in brief, are as follows. In Sec.~\ref{theproposal}, it is proposed that, for any ergodic component of a 2D non-integrable system, the pertinent model for describing the spatial statistical structure of any typical $\Sigma$-type DPP is the two-dimensional homogeneous Poisson point process (henceforth denoted by $\mathbf{P}_2$). In particular, it is argued that the (Euclidean) $k$th-nearest-neighbor distance characteristics ($k=1,2,3,\ldots$) of any such $\Sigma$-type DPP should be consistent with those theoretically predicted for $\mathbf{P}_2$. A concise review of pertinent details concerning $\mathbf{P}_2$ and the associated $k$th-nearest-neighbor distance distributions for $\mathbf{P}_2$ is given in Sec.~\ref{P2Review}. The ideas and arguments put forward in Sec.~\ref{theproposal} are then numerically tested using three exemplary model systems: the cardioid billiard (Sec.~\ref{modexample}), the semi-circular mushroom billiard (Sec.~\ref{mushexample}), and the 2D H\'{e}non-Heiles system (Sec.~\ref{HHexample}). The results of the numerical experiments as well as their significance to classical and quantum chaos are discussed in Sec.~\ref{diskus}, and concluding remarks are given in Sec.~\ref{conc}.

\section{Interpoint Distance Characteristics of $\Sigma$-type DPPs in 2D Non-Integrable Systems: The Ergodic Case}\label{theproposal} 

\begin{figure}
\scalebox{0.651}{\includegraphics*{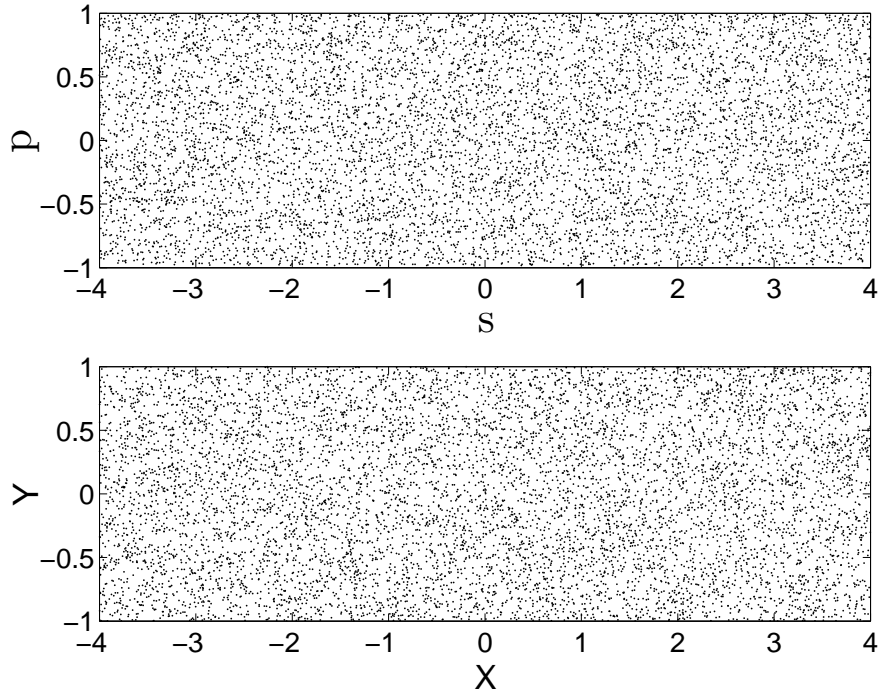}}
\caption{\label{P2P2andWig} (Top) A typical pseudotrajectory of the Poincar\'{e} map of the cardioid billiard. (In this instance, the map was iterated 8000 times.) The phase space of the map is the Poincar\'{e} surface-of-section (SOS) coordinatized using the canonical Birkhoff coordinates $(s,p)$. For the cardioid, the SOS $\Sigma=\{(s,p): s\in[-4,4], p\in[-1,1]\}$. (Bottom) A realization of a two-dimensional homogeneous Poisson point process of intensity $\rho=500$ generated from a binomial point process with $N=8000$ points in the rectangle $[-4, 4] \times [-1, 1]$.}
\end{figure}

In numerical Poincar\'{e} surface-of-section (SOS) computations, the usual procedure is to compute numerical trajectories (i.e., `pseudotrajectories') for a large number of initial conditions and then observe the resulting point pattern that ensues from the intersection of these numerically-computed trajectories with a chosen SOS. In 2D conservative systems, the visual signature of fully developed chaos is an apparently random scatter of points (on the SOS) generated from \emph{one} initial condition. What are the spatial statistical properties of such a point pattern? For illustrative purposes, consider for a moment the point pattern on the SOS shown in the top panel of Fig.~\ref{P2P2andWig}, which was generated by iterating the Poincar\'{e} map of the 2D cardioid billiard (see Sec.~\ref{modexample} for background details). Interestingly, this point pattern (obtained from classical deterministic laws and equations) is \emph{visually} indistinguishable from a realization of a Poisson point process in the SOS (see lower panel of Fig.~\ref{P2P2andWig}). Are the spatial statistical properties of the point pattern obtained from classical mechanics (which is actually a discrete pseudotrajectory of the Poincar\'{e} map) indistinguishable from those of any suitably-defined realization of a two-dimensional Poisson point process? In other words: Are the spatial statistical properties of this $\Sigma$-type DPP consistent with those theoretically predicted for the two-dimensional homogeneous Poisson point process ($\mathbf{P}_2$)? 

Admittedly, there is a certain degree of vagueness in the preceding line of questioning since ``spatial statistical properties" is a rather broad term and can refer to any of a vast number of commonly encountered statistical quantities, relationships, and methodologies based on distance, area, orientation, and/or other geometric descriptors of a point process. For the present, suppose ``spatial statistical properties" refers to any spatial statistical quantity or relationship (pertinent to point processes) that involves only the distances between specified points of a point process. 
The distributions of the $k$th-nearest-neighbor distances ($k=1,2,3,\ldots$), for example, are among the most rudimentary ``distance characteristics''  of a point process and are often used to characterize spatial point patterns that arise from theoretical models and physical data \cite{StoyandStoy,spatstats,StoyandStoy2}. Although no canonical metric exists for measuring distances between arbitrary points in classical phase space, the familiar Euclidean metric can be used without restriction. (Other metrics could be potentially useful depending on the specific system(s) of interest but non-Euclidean metrics will not be given further consideration here.) One approach then to addressing the above-posited questions is to study the $k$th-nearest-neighbor distance distributions ($k$th-NNDDs) of  the points on a given pseudotrajectory of the flow that intersect a chosen SOS, or equivalently \cite{RemPoinMap}, to study the distributions of the $k$th-nearest-neighbor distances between the points of a given `discrete time' pseudotrajectory of the 2D Poincar\'{e} map. 

For a fully chaotic 2D system, almost any pseudotrajectory of the Poincar\'{e} map will explore the full phase space of the map ergodically, i.e., almost any pseudotrajectory of the Poincar\'{e} map will densely and uniformly cover the entire SOS. The $k$th-nearest-neighbor distance distributions of such a pseudotrajectory should (due to denseness and uniformity) be well modeled by the corresponding distributions theoretically predicted for $\mathbf{P}_2$. The same should be true of any chaotic pseudotrajectory that ergodically explores any positive-measure subset of the total available phase space. (In other words, individual chaotic trajectories need not densely cover the entire phase space.)  Successive point intersections of any such pseudotrajectory with a canonical SOS will (after sufficient time) uniformly cover some subset $W$ of the SOS. The ensuing point set should be indistinguishable (insofar as its distance characteristics are concerned) from any suitably defined realization of a Poisson point process in $W$ (of appropriate intensity). Even in the case of generic mixed systems where no ergodic components exist (i.e., no positive-measure regions of phase space are completely devoid of islands), there generally exist conditions under which the chaotic pseudotrajectories explore, \emph{nearly uniformly}, most of the available phase space. In such cases, typical pseudotrajectories should still possess $k$th-NNDDs that are reasonably well modeled by the theoretical $k$th-NNDDs for $\mathbf{P}_2$ (but likely less accurately than in the previously discussed ergodic cases). The preceding claims naturally require verification and indeed the intent in the following sections of the paper is to validate these claims numerically. Before proceeding, the pertinent details concerning the $k$th-NNDDs for $\mathbf{P}_2$ are briefly reviewed. 

\section{Unit-Mean $k$th-Nearest-Neighbor Distance Distributions for $\mathbf{P}_2$}\label{P2Review}

The homogeneous Poisson point process in $\mathbbm{R}^2$ (denoted by $\mathbf{P}_2$) is essentially the limit of a simpler stochastic model: the binomial point process in $\mathbbm{R}^2$. The latter model consists of $N$ random points uniformly distributed in a compact subset $W$ of $\mathbbm{R}^2$. If the area bounded by $W$ is $A(W)$ and we take the limits $N\rightarrow\infty$ and $A(W)\rightarrow\infty$ in such a way that $N/A(W)\equiv\rho$ remains constant, then the limiting stochastic point process is $\mathbf{P}_2$ (with intensity $\rho$). As a conceptual example, suppose $W$ is a rectangle with side lengths $L$ and $H$. If we take the limits $N\rightarrow\infty$, $L\rightarrow\infty$, and $H\rightarrow\infty$ in such a way that $N/LH\equiv\rho$ remains constant, then the limiting point process is $\mathbf{P}_2$ (with intensity $\rho$). More precise and technical definitions of $\mathbf{P}_2$ can be found in the mathematical literature \cite{StoyandStoy,SKM}, but the technicalities involved are not relevant to the following developments. 
 
The $k$th-nearest-neighbor distance distribution ($k$th-NNDD) for $\mathbf{P}_2$, denoted here by $\mathrm{D}(s;k)$, gives the probability $\mathrm{D}(s;k)\d s$ of finding the $k$th-nearest neighbor to a given point of $\mathbf{P}_2$ at a distance between $s$ and $s+\d s$. 
It can be shown that the $k$th-NNDD for $\mathbf{P}_2$ is \cite{StoyandStoy} 
\be\label{KNNDDR2}
\mathrm{D}(s;k)={2\left(\rho\pi\right)^k \over \Gamma(k)} s^{2k-1}
\exp\left(-\rho\pi s^2\right).
\ee
It is easy to verify that the above distribution is normalized (i.e., $\int_0^\infty \mathrm{D}(s;k)\d s=1$) and that the mean $k$th-nearest-neighbor distance is 
\be\label{sbarPPPK}
\bar{s}=\int_0^\infty s\mathrm{D}(s;k)\d
s={\Gamma\left(k+{1\over2}\right)\over\Gamma(k)\sqrt{\rho\pi}},
\ee
where $\Gamma(x)$ is the standard Gamma function. If we transform to the random variable $S=s/\bar{s}$, the distribution (\ref{KNNDDR2}) becomes
\begin{subequations}\label{KNNDDRMT}
\be
\mathrm{D}(S;k)={2\alpha^k\over\Gamma(k)}S^{2k-1}\exp
\left(-\alpha S^2\right),
\label{KNNDDRMTp1}
\ee
where
\be
\alpha=\left[{{\Gamma\left(k+{1\over2}\right)}\over\Gamma(k)}\right]^2.
\label{KNNDDRMTp2}
\ee
\end{subequations} 
Note that the distribution (\ref{KNNDDRMT}) is also normalized (i.e., $\int_0^\infty \mathrm{D}(S; k)\d S=1$), has unit mean (i.e., $\int_0^\infty S \mathrm{D}(S; k)\d S=1$), and most importantly, is intensity-\emph{independent} (i.e., does not explicitly depend on $\rho$). For future reference, the \emph{unit-mean} $k$th-NNDDs for $k=1,2$, and $3$ are:  
\be\label{WigdisRMT}
\mathrm{D}(S;1)={\pi\over2}S\exp\left(-{\pi\over4}S^2\right), 
\ee
\be\label{WigGinib}
\mathrm{D}(S;2)={3^4\pi^2\over2^7}S^3
\exp\left(-{3^2\pi\over2^4}S^2\right), 
\ee 
and   
\be\label{P2Keq3}
\mathrm{D}(S;3)={15^6\pi^3\over2^{24}}S^5\exp\left(-{15^2\pi\over2^8}S^2\right).
\ee
The distributions (\ref{WigdisRMT}) and (\ref{WigGinib}), well known to practitioners of random matrix theory, are the Wigner and Ginibre distributions, respectively \cite{Mehta,Haake,meandjohn}. 

\section{Example 1: 2D Cardioid Billiard}\label{modexample} 

\begin{figure}
\scalebox{0.397}{\includegraphics*{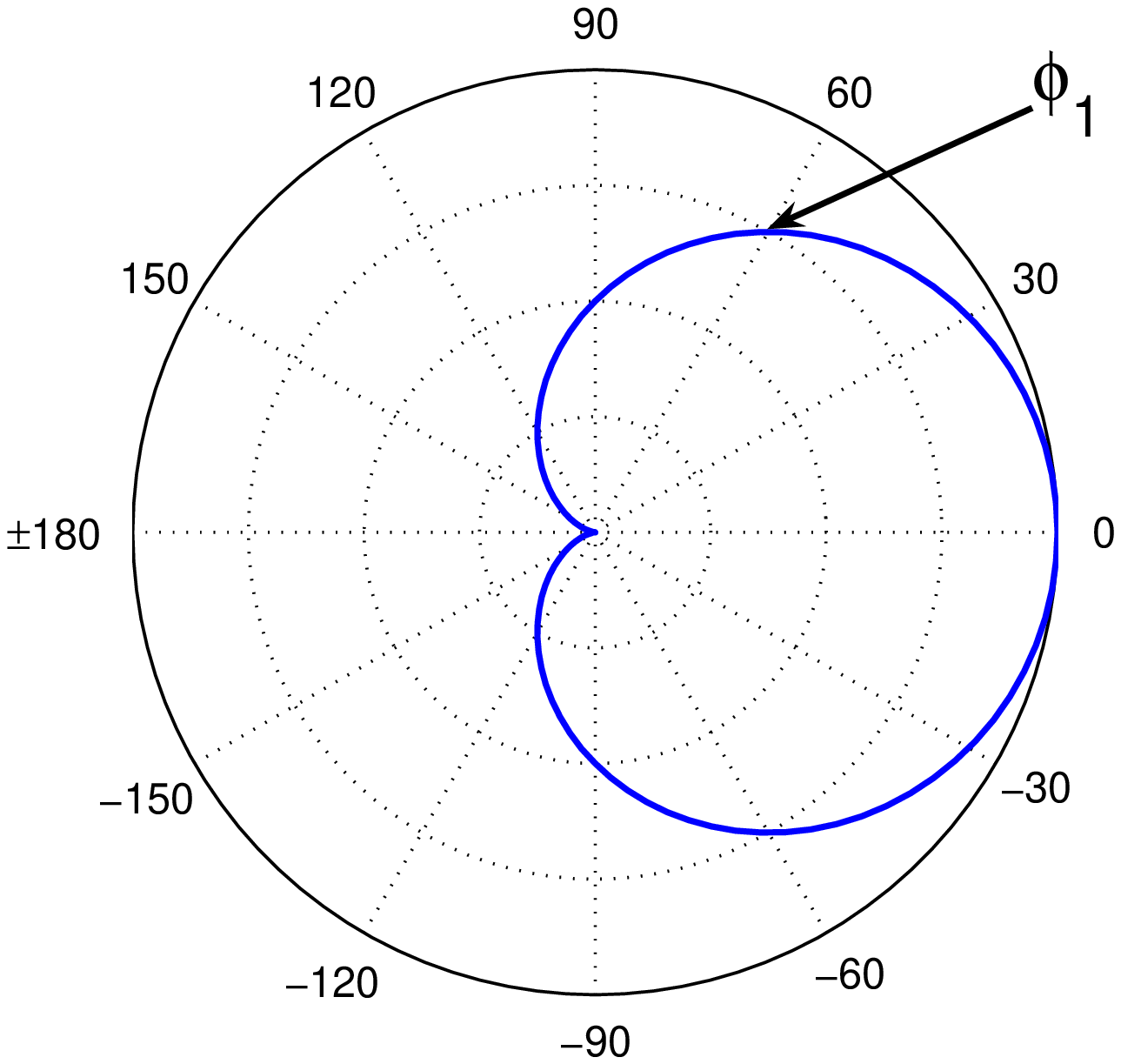}}
\hspace*{1cm}
\scalebox{0.451}{\includegraphics*{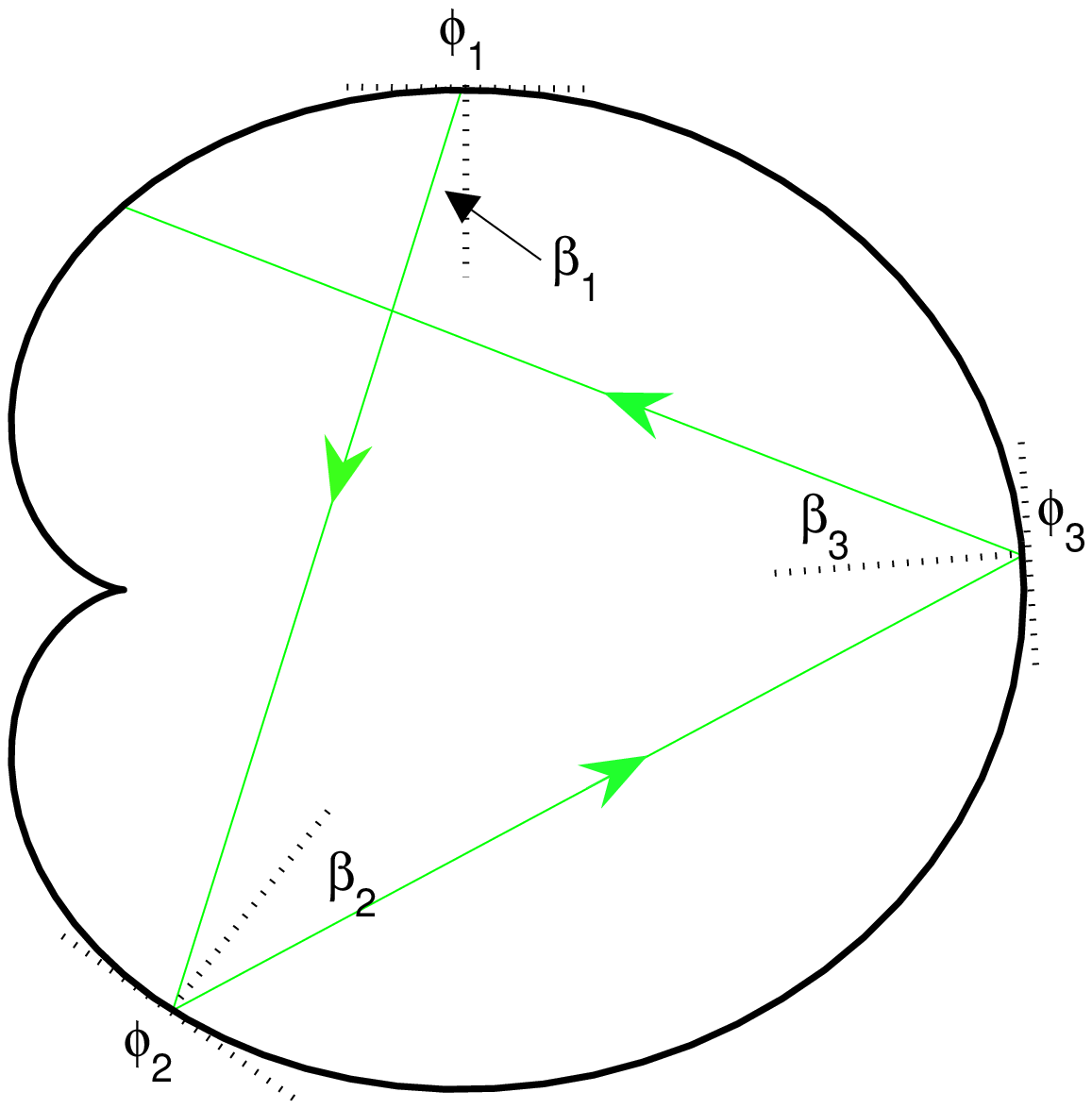}}
\caption{\label{coords} (Left) The two-dimensional cardioid billiard
  with polar axes included for reference. (Right) The first segments of 
  a typical trajectory specified by a set of polar angles $\phi$ and
bounce angles $\beta$. In this instance, $\phi_1=\pi/3$ and $\beta_1=\pi/12$.}
\end{figure}

The first of the three claims put forward in Sec.~\ref{theproposal} can be expressed as follows: \emph{For a fully chaotic 2D Hamiltonian system, typical pseudotrajectories of the Poincar\'{e} map possess (Euclidean) $k$th-nearest-neighbor distance characteristics consistent with those theoretically predicted for $\mathbf{P}_2$}. There are many well-known examples of 2D fully chaotic systems, any of which could serve to validate (or invalidate) the preceding statement (henceforth referred to as Claim 1). Among the most celebrated examples are planar hyperbolic billiards such as the cardioid billiard shown in Fig.~\ref{coords}. In the following, the validity of Claim 1 is demonstrated in the representative case of the cardioid billiard. 

\begin{figure}
\scalebox{0.531}{\includegraphics*{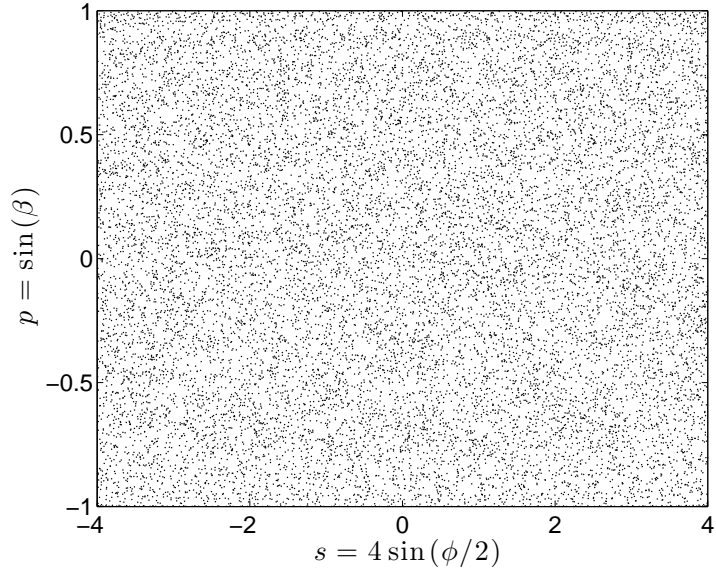}}
\caption{\label{Pnesection} A typical pseudotrajectory of the Poincar\'{e} map of the cardioid billiard. In this instance, the pseudotrajectory was launched with initial condition $(s_1=4\sin(\pi/6),p_1=\sin(\pi/12))$ and the map iterated 16000 times.}
\end{figure}

Consider then the motion of a point particle confined in a cardioid billiard. 
In order to describe the particle's motion, it is sufficient to know the points of reflection on the boundary and the corresponding direction afterwards. Particle trajectories in the
billiard can be fully specified using two angles, the polar angle
$\phi$, which determines the position of the bounce point on the boundary, 
and the bounce angle $\beta$, which is the angle 
between the outgoing ray and the local 
normal vector pointing into the billiard. Individual trajectories can
then be given as a set of angle pairs: 
$\{(\phi_1,\beta_1),(\phi_2,\beta_2),\dots,(\phi_n,\beta_n),\ldots\}$ 
(see Fig.~\ref{coords}). This description of the particle motion leads directly to the `billiard map' (i.e., the Poincar\'{e} map for the billiard flow). For a billiard flow, the most convenient SOS is the set of collision points of the flow. Any trajectory of the billiard flow intersects this SOS whenever it reflects at the boundary. This defines the standard billiard map, and the SOS defined in this way is the phase space of that map. The standard coordinatization of such a SOS employs the canonical Birkhoff coordinates $(s,p)$, where
$s=4\sin(\phi/2)$ is the arclength position of a bounce measured along the
boundary, and $p=\sin(\beta)$ is the tangential momentum, that is, the
momentum component parallel to the boundary at the bounce point \cite{standardMandV}. 

For the cardioid, the Poincar\'{e} section $\Sigma=\{(s,p): s\in[-4,4], p\in[-1,1]\}$. 
The Poincar\'{e} map $\mathcal{B}$ specifies completely the evolution of position 
and momentum from one collision with the boundary to the next. The map $\mathcal{B}$ (sometimes also called the Birkhoff map)
is obtained from calculating the image point $\xi'=(s',p')\in\Sigma$ 
of a given point $\xi=(s,p)\in\Sigma$ (i.e., $\mathcal{B}: \Sigma \rightarrow \Sigma$, $\xi=(s,p) \longmapsto \mathcal{B}(s,p)=(s',p')=\xi'$). Details concerning the Poincar\'{e} map for the cardioid billiard can be found in Ref.~\cite{BackandcoJPA}. 

In studying `typical' pseudotrajectories of $\mathcal{B}$, the set of initial conditions which start at or will immediately hit the cusp (the one singular point of the billiard) is excluded, as is the set of tangential collision points (the so-called ``fixed points" of $\mathcal{B}$) that result in a sliding motion where the moving point particle slides along the boundary wall. These sets are of measure zero, and thus almost all initial conditions produce long pseudotrajectories suitable for analysis. A typical pseudotrajectory of $\mathcal{B}$ is shown in Fig.~\ref{Pnesection}. Using the initial condition $(s_1,p_1)=(4\sin(\phi_1/2),\sin(\beta_1))=(4\sin(\pi/6),\sin(\pi/12))$, the Poincar\'{e} map $\mathcal{B}$ was iterated $16000$ times. The resulting pseudotrajectory 
is shown plotted in the phase space of $\mathcal{B}$ (i.e., the SOS). The task now is to analyze the distances between the points of this pseudotrajectory. 

The distance between two points $\xi_i=(s_i,p_i)$ and $\xi_j=(s_j,p_j)$ on the SOS 
$\Sigma$ is defined using the usual Euclidean metric: 
$\Delta(\xi_i,\xi_j)=\sqrt{(s_i-s_j)^2+(p_i-p_j)^2}$. 
The distance between a given point $\xi_i$ and its nearest neighbor  
is then defined by 
$d^{(1)}_{i} = \text{min} \left\{\Delta\left(\xi_i,\xi_j\right) 
: i,j=1,\ldots,N~(j \neq i)\right\}$, and similarly the distance between 
$\xi_i$ and its furthest neighbor is defined by 
$d^{(N)}_{i} = \text{max} \left\{\Delta\left(\xi_i,\xi_j\right) 
: i,j=1,\ldots,N~(j \neq i)\right\}$. If, for a given point  
$\xi_i$, the distances $\left\{\Delta\left(\xi_i,\xi_j\right) 
: i,j=1,\ldots,N~(j \neq i)\right\}$ are sorted by size (in ascending order), then the 
$k$th-nearest-neighbor distance is the $k$th element of the set
$\{d^{(1)}_i,d^{(2)}_i,\ldots,d^{(k)}_i,\ldots,d^{(N)}_i\}$. 
The set $\{d^{(k)}_i:i=1,\ldots, N\}$ thus defines the (experimental) set of $k$th-nearest-neighbor distances and density histograms of these interpoint distances could (barring some details) be compared with model predictions for $\mathbf{P}_2$ [i.e., Eq.~(\ref{KNNDDR2})]. A more elegant approach is to remove the inherent dependence on the intensity by constructing density histograms of the scaled (and dimensionless) distances $S^{(k)}_i = d^{(k)}_i / \bar{d}^{(k)}$, where $\bar{d}^{(k)}=(1/N)\sum_{i=1}^N d^{(k)}_i$, is the 
mean $k$th-nearest-neighbor distance. Doing so generates (for a given value of $k$) an experimental density histogram that can then be directly compared to the corresponding intensity-\emph{independent} $k$th-NNDD given in Sec.~\ref{P2Review} [i.e., Eq.~(\ref{KNNDDRMT})]. 

\begin{figure}
\scalebox{0.511}{\includegraphics*{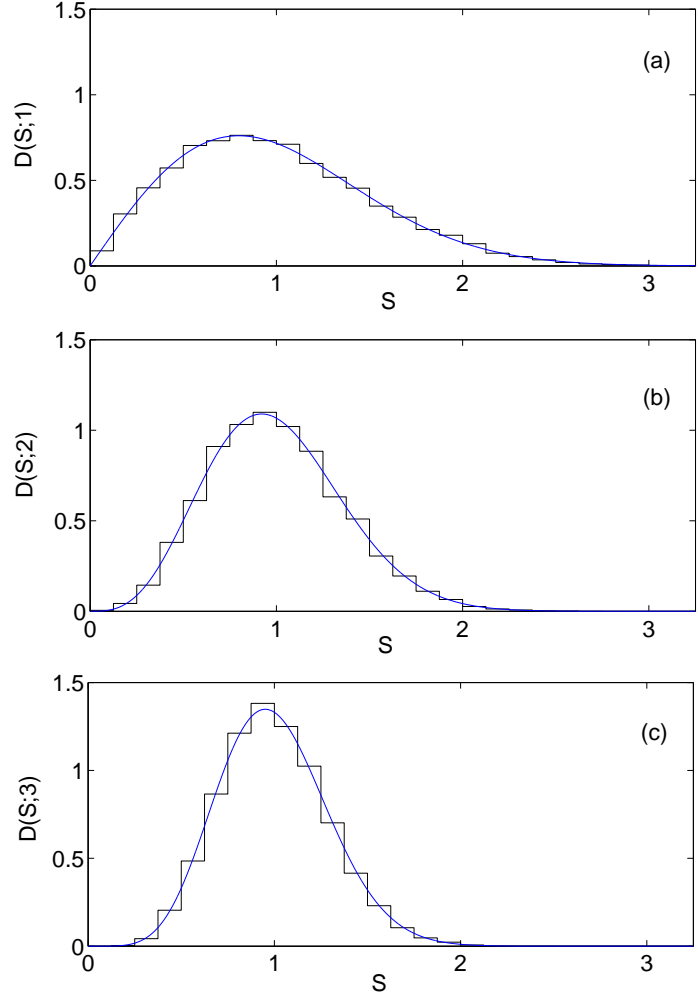}}
\caption{\label{WandG} Near-neighbor distance distributions for the pseudotrajectory shown in Fig.~\ref{Pnesection}. (a) Density histogram of the (scaled) \emph{nearest}-neighbor distances; 
the smooth curve is the Wigner distribution [Eq.~(\ref{WigdisRMT})]. (b) Density histogram of the (scaled) \emph{second}-nearest-neighbor distances; the smooth curve is the Ginibre distribution [Eq.~(\ref{WigGinib})]. (c) Density histogram of the (scaled) \emph{third}-nearest-neighbor distances; the smooth curve is the distribution $\mathrm{D}(S;k=3)$ [Eq.~(\ref{P2Keq3})].}
\end{figure}

For the pseudotrajectory shown in Fig.~\ref{Pnesection}, the density histogram of the (scaled) \emph{nearest}-neighbor distances     
is shown in Fig.~\ref{WandG}(a) and is clearly in accord with the Wigner distribution [Eq.~(\ref{WigdisRMT})]. 
The density histograms of the (scaled) \emph{second}- and \emph{third}-nearest-neighbor distances are shown in 
Figs.~\ref{WandG}(b) and ~\ref{WandG}(c), respectively. The correspondence with the Ginibre distribution [Eq.~(\ref{WigGinib})] in the former case and with the distribution $\mathrm{D}(S;k=3)$ [Eq.~(\ref{P2Keq3})] in the latter is evident.

Similar results were found for several other pseudotrajectories of the same length (six in total were analyzed). Thus, for the cardioid billiard, typical pseudotrajectories of the Poincar\'{e} map have $k$th-nearest-neighbor distance characteristics consistent with those theoretically predicted for $\mathbf{P}_2$. 

Equivalent analyses of some typical pseudotrajectories of the stadium billiard (omitted here) also yield similar results \cite{meunpublished}. These results corroborate the validity of Claim 1. 

\section{Example 2: 2D Semi-Circular Mushroom Billiard}\label{mushexample} 

\begin{figure}
\scalebox{0.423}{\includegraphics*{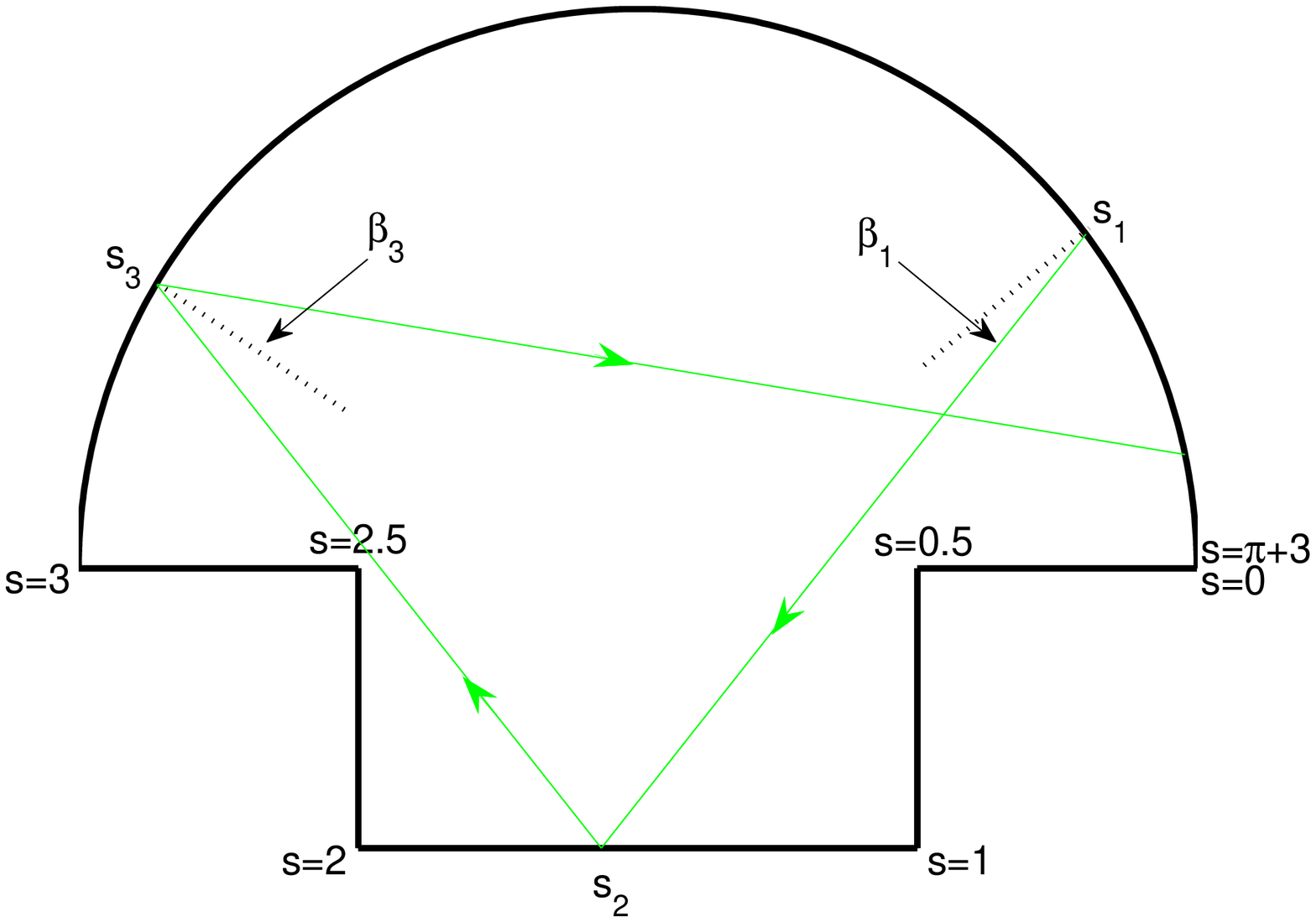}}
\caption{\label{MushDiagram} A simple two-dimensional mushroom billiard composed of a semi-circular cap and rectangular stem. The cap has unit radius, the stem has unit width, and the stem height is half of the stem width. Also shown are the first three segments of a typical trajectory, which is here specified by a set of coordinates $(s,\beta)$, where $s\in[0,\pi+3)$ is the arclength position coordinate of a collision point measured along the billiard's boundary (increasing from zero at the right-most point of the mushroom as the boundary is traversed clockwise) and $\beta$ is the reflection angle at each collision.}
\end{figure}

\begin{figure}
\scalebox{0.551}{\includegraphics*{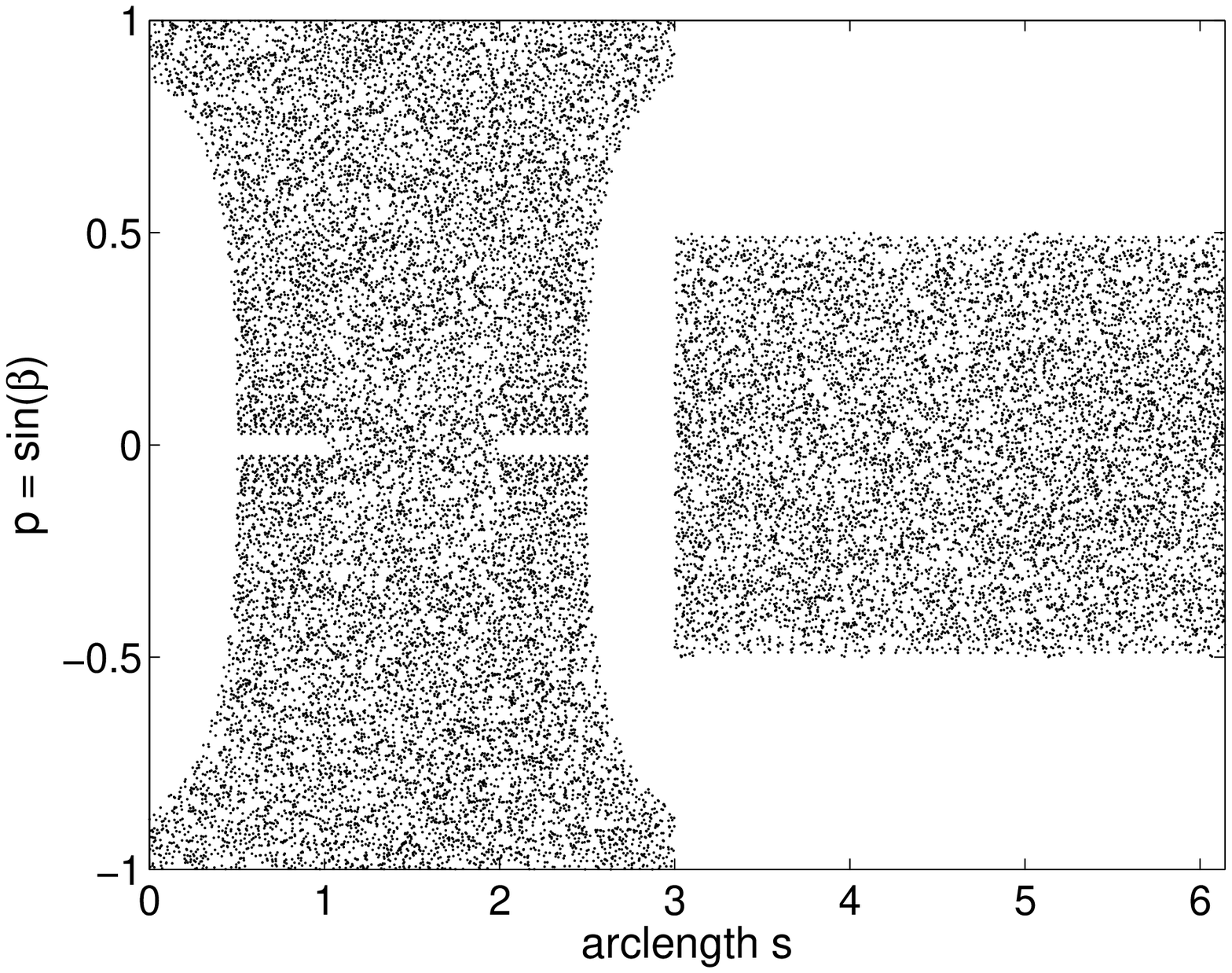}}
\caption{\label{MushSection} A typical chaotic pseudotrajectory of the Poincar\'{e} map of the 2D semi-circular mushroom billiard (see Fig.~\ref{MushDiagram}). The phase space of the map (i.e., the surface-of-section) is coordinatized using the Birkhoff coordinates $(s,p=\sin(\beta))$. In this instance, the pseudotrajectory was evolved from an initial point $(s_1=5.501,p_1=\sin(\pi/12))$ in the ergodic region, and the Poincar\'{e} map was iterated 25000 times.}
\end{figure}

Chaotic trajectories need not necessarily cover densely all of the available phase space in order for their point crossings with a SOS to exhibit the properties shown in Fig.~\ref{WandG}. Trajectories evolving ergodically in any positive-measure subset of the full phase space should have the same spatial statistical properties. Such dynamical evolution is known to occur but is relatively rare among non-integrable Hamiltonian systems. The most well-known examples of mixed systems with a so-called sharply-divided phase space are the Bunimovich mushroom billiards \cite{BuniMushroom}. The mushroom billiard shown in Fig.~\ref{MushDiagram} having a semi-circular cap and rectangular stem is the simplest example of this special class of mixed systems. The phase space of this mixed system is unusual in that it has a single regular region and a single chaotic region, and no bordering Kolmogorov-Arnold-Moser (KAM) hierarchy \cite{BuniMushroom}. (In generic mixed systems, the border between a regular and chaotic region manifests a complex hierarchical structure of KAM islands and cantori.) Interestingly, mushrooms can be designed so as to have any desired number of positive-measure ergodic components and any number of islands of stability. Furthermore, one can controllably alter the relative volume fractions in phase space containing regular and chaotic trajectories by simply varying the dimensions of the mushroom. Aside from these particulars, the main point is that the mushroom shown in Fig.~\ref{MushDiagram} has an ergodic region (i.e., there exists a positive-measure subset of the full phase space wherein the dynamics is ergodic). As argued in Sec.~\ref{theproposal}, the $k$th-nearest-neighbor distance characteristics of any typical $\Sigma$-type DPP initiated in the ergodic region should be consistent with model predictions for $\mathbf{P}_2$. The validity of the preceding claim is demonstrated below.

As in the case of the cardioid billiard, the SOS is coordinatized using the Birkhoff coordinates $(s,p)$, where $s$ is the arclength position coordinate of a collision point (measured along the boundary starting from zero at the right-most point of the mushroom and increasing to its maximum value as the boundary is traversed clockwise), and $p=\sin(\beta)$ is the momentum component parallel to the boundary at the collision point (see Fig.~\ref{MushDiagram}). For the mushroom billiard shown in Fig.~\ref{MushDiagram}, the SOS $\Sigma=\{(s,p): s\in[0,\pi+3), p\in[-1,1]\}$. Recall from Sec.~\ref{modexample} that the SOS is the phase space of the Poincar\'{e} map and that the conventional Poincar\'{e} map $\mathcal{B}$ for a billiard flow specifies the evolution of position and momentum from one collision with the boundary to the next: $\mathcal{B}: \Sigma \rightarrow \Sigma$, $\xi=(s,p) \longmapsto \mathcal{B}(s,p)=(s',p')=\xi'$. A typical chaotic pseudotrajectory of $\mathcal{B}$ is shown in Fig.~\ref{MushSection}. This particular pseudotrajectory was generated by iterating the Poincar\'{e} map $\mathcal{B}$ of the mushroom billiard $25000$ times starting from the initial condition $(s_1,p_1)=(5.501,\sin(\pi/12))$, which lies in the ergodic domain \cite{singaset}. The procedure for computing the $k$th-nearest-neighbor distances between the points of this pseudotrajectory is the same as the one detailed in Sec.~\ref{modexample}. 

\begin{figure}
\scalebox{0.511}{\includegraphics*{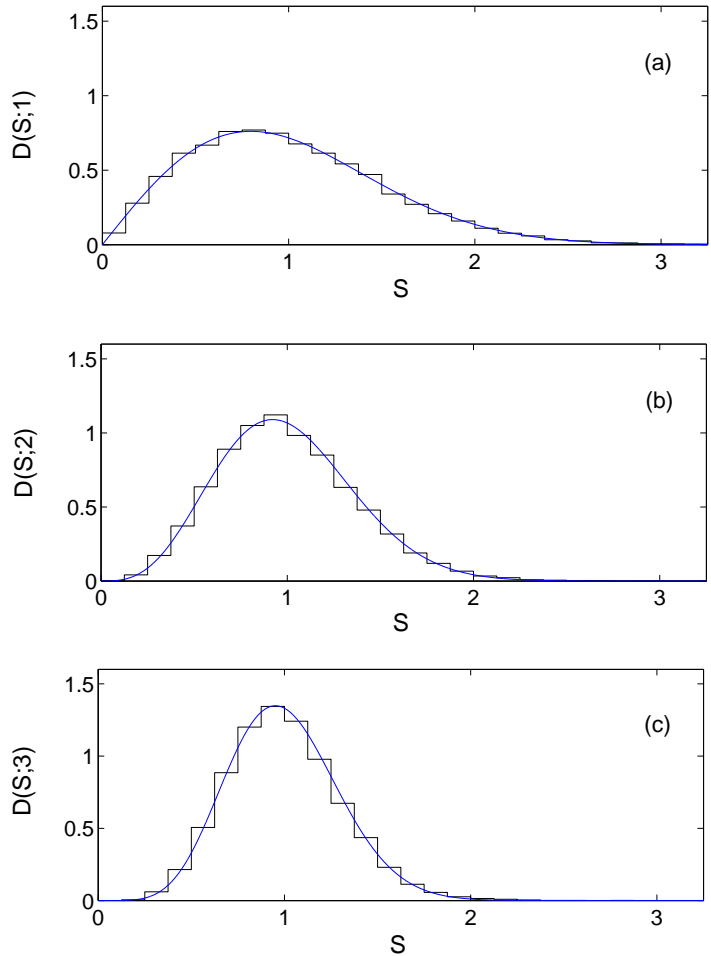}}
\caption{\label{MushHistFinal} Near-neighbor distance distributions for the pseudotrajectory shown in Fig.~\ref{MushSection}. (a) Density histogram of the (scaled) \emph{nearest}-neighbor distances; 
the smooth curve is the Wigner distribution [Eq.~(\ref{WigdisRMT})]. (b) Density histogram of the (scaled) \emph{second}-nearest-neighbor distances; the smooth curve is the Ginibre distribution [Eq.~(\ref{WigGinib})]. (c) Density histogram of the (scaled) \emph{third}-nearest-neighbor distances; the smooth curve is the distribution $\mathrm{D}(S;k=3)$ [Eq.~(\ref{P2Keq3})].}
\end{figure}

Density histograms of the $k$th-nearest-neighbor distances (for $k=1,2$, and $3$) are shown in Fig.~\ref{MushHistFinal}. In each case, the sample histogram is consistent with model predictions for $\mathbf{P}_2$ [see Eqs.~(\ref{WigdisRMT}), (\ref{WigGinib}), and (\ref{P2Keq3}) of Sec.~\ref{P2Review}]. Similar results were found for several other pseudotrajectories of the similar length. Thus, in the case of this simple mushroom billiard (which has a sharply-divided phase space with one ergodic region), typical pseudotrajectories of $\mathcal{B}$ from the ergodic region of the SOS possess $k$th-nearest-neighbor distance characteristics consistent with those of $\mathbf{P}_2$. 

\section{Example 3: 2D H\'{e}non-Heiles Potential}\label{HHexample} 

Corresponding analyses of $\Sigma$-type DPPs in generic mixed systems lie outside the scope of the present paper. There are however two important limiting cases where typical $\Sigma$-type DPPs can be expected to have properties similar to those observed in the prior two examples: (i) generic systems above the energy threshold; and (ii) KAM-type systems (i.e., Hamiltonian systems to which the KAM theorem applies) above the stochasticity threshold. An explicit example of the former is considered in this section.

In generic mixed systems, there are no ergodic components (i.e., no positive-measure regions of phase space completely devoid of islands). Nevertheless, there generally exist values of a system parameter at which there are only a few observable islands and the fraction of the phase space volume occupied by these islands is small (e.g., $<0.1$). In such cases, the chaotic sea covers \emph{nearly uniformly} most of the available phase space, and thus by extension, the $k$th-nearest-neighbor distance characteristics of any typical $\Sigma$-type DPP initiated in the chaotic sea should be very similar to those that would be observed for a typical $\Sigma$-type DPP evolved in a strictly ergodic region.

The system parameter referred to above could be a coupling or perturbation parameter (e.g., kicked rotor \cite{Chirikov79}), a shape parameter defining a family of billiards (e.g., the $\delta$ parameter defining the family of lemon billiards \cite{Lemon}), or even simply the total system energy in time-independent potential systems (e.g., H\'{e}non-Heiles \cite{HH64}, Pullen-Edmonds \cite{EPPOT} etc.). The typical scenario in the last case (the most fundamental case for smooth Hamiltonian systems) is a phase space dominated by regular trajectories at low energies and chaotic trajectories at high energies with a mix of regular and chaotic trajectories at intermediate values of the energy. In the well-known H\'{e}non-Heiles system, for example, most trajectories are quasi-periodic at $E=1/12$ and chaotic at $E=1/6$ with a mix of regular and chaotic trajectories at $E=1/8$ (see Ref.~\cite{HH64}). The claim in this case then is that, at the critical energy $E=1/6$, the interpoint distance characteristics of any typical $\Sigma$-type DPP should be adequately modeled by those of $\mathbf{P}_2$. In particular, good agreement between the sample $k$th-NNDDs of the DPP and the theoretical $k$th-NNDDs predicted for $\mathbf{P}_2$ should be observed. In the following, this claim is numerically confirmed. 

\begin{figure}
\scalebox{0.553}{\includegraphics*{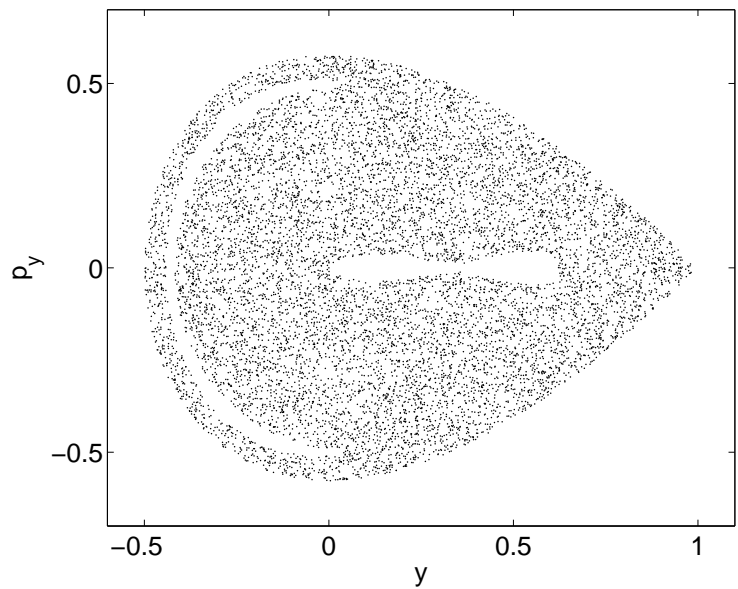}}
\caption{\label{HHSection} Successive intersections of a typical pseudotrajectory of (\ref{EOMHH}) with the $x=0$ plane in the upward direction ($p_x>0$). The pseudotrajectory was evolved for $70000$ ``seconds'' from an initial point $(x(0)=0,p_x(0)=\sqrt{1/3},y(0)=0,p_y(0)=0)$ in the $E=1/6$ energy shell and intersects the Poincar\'{e} plane (in the upward direction) $10313$ times during its evolution.}
\end{figure}

\begin{figure}
\scalebox{0.557}{\includegraphics*{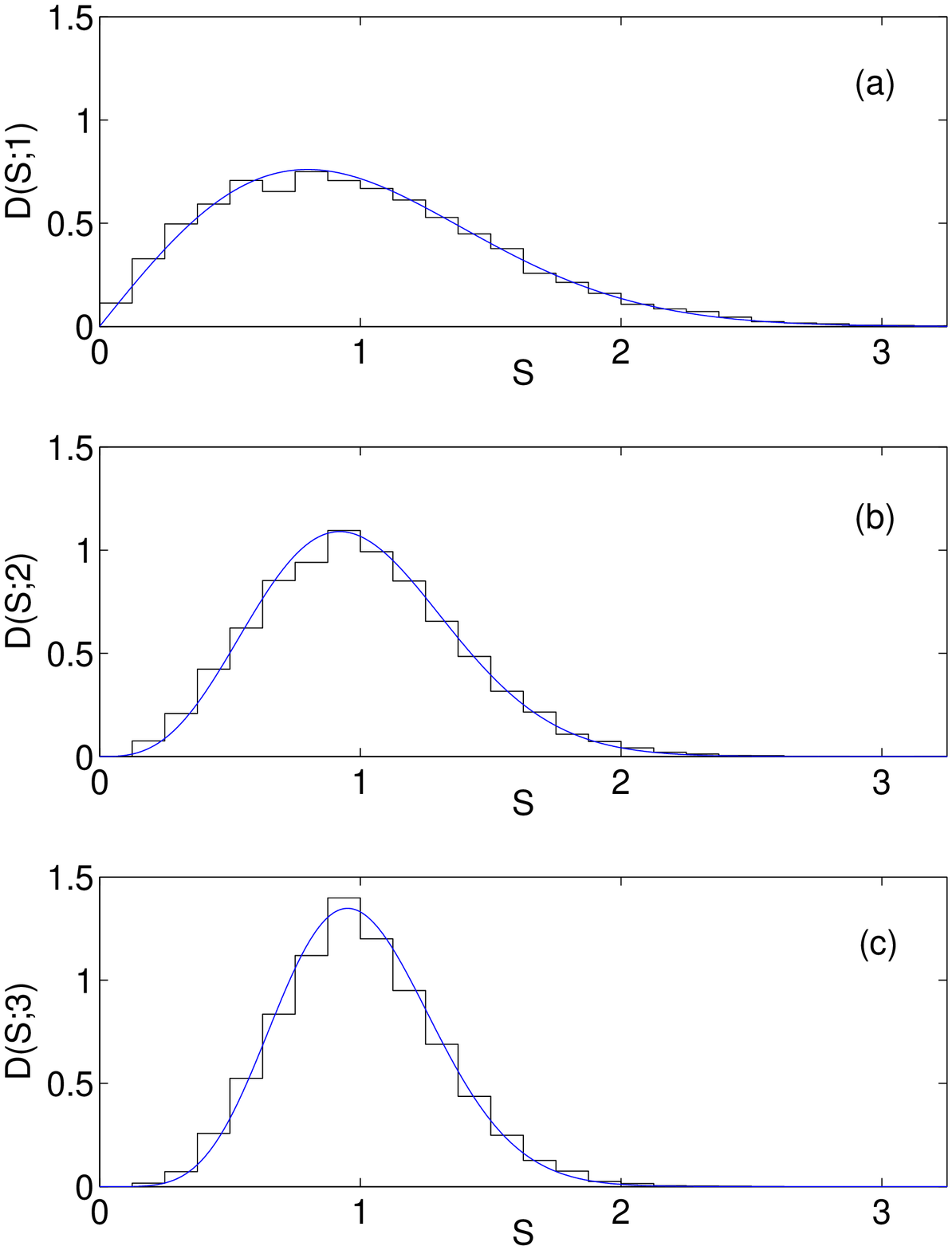}}
\caption{\label{HHHistFinal1} Near-neighbor distance distributions for the $\Sigma$-type DPP shown in Fig.~\ref{HHSection}. (a) Density histogram of the (scaled) \emph{nearest}-neighbor distances; the smooth curve is the Wigner distribution [Eq.~(\ref{WigdisRMT})]. (b) Density histogram of the (scaled) \emph{second}-nearest-neighbor distances; the smooth curve is the Ginibre distribution [Eq.~(\ref{WigGinib})]. (c) Density histogram of the (scaled) \emph{third}-nearest-neighbor distances; the smooth curve is the distribution $\mathrm{D}(S;k=3)$ [Eq.~(\ref{P2Keq3})].}
\end{figure}

The motion of a point particle of unit mass in a smooth 2D time-independent potential $V(x,y)$ is governed by the Hamilton equations of motion 
\be\label{HEOMG}
\dot{x}=p_x, ~~~\dot{p}_x=-{\partial V\over\partial x}, ~~~\dot{y}=p_y, ~~~\dot{p}_y=-{\partial V\over\partial y}. 
\ee
In the case of the 2D H\'{e}non-Heiles potential  \cite{HH64}
\be\label{HHP}
V(x,y)={1\over2}\left(x^2+y^2\right)+x^2y-{1\over3}y^3,
\ee
the system (\ref{HEOMG}) becomes
\begin{subequations}\label{EOMHH}
\begin{eqnarray} 
\dot{x}&=&p_x, \\
\dot{p}_x&=&-(x+2xy), \\
\dot{y}&=&p_y, \\
\dot{p}_y&=&-(y+x^2-y^2).
\end{eqnarray}
\end{subequations}

The most common and convenient Poincar\'{e} surfaces for a system such as (\ref{HHP}) are planes with either $x=0$ or $y=0$. The former will be employed here. Successive intersections of a pseudotrajectory with the plane $x=0$ can occur in two different directions depending on the sign of $\dot{x}$ when the pseudotrajectory crosses the plane. The Poincar\'{e} map can be fully specified by considering only crossings in the positive direction (i.e., $\dot{x}>0$). The canonical SOS to be employed here is thus defined by $x=0$ and $\dot{x}>0$. Since $y$ and $\dot{y}\equiv p_y$ remain arbitrary, the SOS is thus coordinatized by $(y,\dot{y})$. 

A simple procedure for generating the DPPs on the SOS (i.e., the pseudotrajectories of the Poincar\'{e} map) was employed here which can be briefly summarized as follows. First, the energy was fixed at $E=1/6$ and initial conditions of the form $(x(0)=0,p_x(0),y(0)=0,p_y(0))$ were specified subject to the condition $p_x^2(0)+p_y^2(0)=2E$. The equations of motion (\ref{EOMHH}) were then solved numerically in \texttt{MATLAB} using the ODE solver \texttt{ode45}, which utilizes fourth- and fifth-order Runge-Kutta formulas \cite{numdet1}. Finally, the event location property of the ODE solver was employed to determine when $x(t)=0$ and $p_x(t)>0$ occur simultaneously and to evaluate the solution components $y(t)$ and $p_y(t)$ at these instances. 

A typical chaotic pseudotrajectory of the Poincar\'{e} map is shown in Fig.~\ref{HHSection} (see caption of Fig.~\ref{HHSection} for more details). Density histograms of the $k$th-nearest-neighbor distances (again computed using the procedure detailed in Sec.~\ref{modexample}) for $k=1,2$, and $3$ are shown in Fig.~\ref{HHHistFinal1}. Even for this relatively short pseudotrajectory (the pseudotrajectory analyzed in the previous example was roughly two and a half times longer) the overall agreement with model predictions for $\mathbf{P}_2$ is quite satisfactory. Similar results were found for several other pseudotrajectories of similar length. While the expectation is that substantially longer and more accurate pseudotrajectories will yield better agreement with $\mathbf{P}_2$ predictions than those obtained from the simple numerical techniques employed here, the present results already substantiate the general claim, namely, that for a 2D generic mixed system above the energy or stochasticity threshold, the Euclidean $k$th-nearest-neighbor distance characteristics of any typical $\Sigma$-type DPP are well modeled by the corresponding characteristics theoretically predicted for $\mathbf{P}_2$. 

\section{Discussion}\label{diskus}

\subsection{Zaslavsky's Dictum on Chaotic Trajectories and Stochastic Processes}

To contextualize the numerical results of Secs.~\ref{modexample} - \ref{HHexample}, it is useful to quote here a certain remark (pertaining to chaotic particle trajectories in generic Hamiltonian systems) from a review paper by Zaslavsky \cite{Zassy2002}: \textit{``Trajectories, being considered as a kind of stochastic process, do not behave like well-known Gaussian, Poissonian, Weiner, or other processes."} While Zaslavsky's remark may be true for chaotic pseudotrajectories evolving in a generically mixed phase space, chaotic pseudotrajectories evolving in an ergodic region do in fact behave like one of the well-known Poissonian processes. In the latter case, and in the specific context of 2D non-integrable systems, chaotic pseudotrajectories of the Poincar\'{e} map, ``being considered as a kind of stochastic process'' (to use Zaslavsky's phrasing), behave like a 2D Poisson point process. Under certain conditions, the preceding holds (at least approximately) even in generic mixed systems, as demonstrated in Sec.~\ref{HHexample}. 

\subsection{Exact Trajectories and Shadowing}

The numerical results describe the $k$th-nearest-neighbor distance characteristics of \emph{pseudo}trajectories. This begs the question: Do the same results hold for the \emph{exact} trajectories of a Poincar\'{e} map? This is a question that is difficult to answer definitively without detailed analysis. Chaotic pseudotrajectories emulate the true dynamics of a given system only when `shadowed' by exact trajectories of the system, and are otherwise (individually) meaningless. Shadowing of numerical trajectories is a fundamental issue, in particular, for strictly non-hyperbolic chaotic systems. In hyperbolic systems (i.e., `hard chaotic' systems), which is an extreme and rather exceptional case, the existence of shadowing trajectories for all pseudotrajectories is guaranteed (see, for example, Ref.~\cite{katok}). For such systems, conclusions about the statistical properties of numerical trajectories will also generally apply to exact trajectories. Long shadowing trajectories are not precluded for all non-hyperbolic chaotic systems (for example, the shadowing property has been established for the standard map \cite{SM1}), but their consideration introduces technical questions about how accurate and for how long numerical trajectories are valid. 

The shadowing property also holds for pseudotrajectories of chaotic billiards, under certain conditions \cite{CM2006}. (``Chaotic billiards" includes all the famous examples, such as the stadium billiard, the semi-dispersing Sinai billiard, the diamond billiards, and many others.) In applying shadowing arguments to chaotic billiards, it is necessary to exclude the vicinity of singularities (i.e., discontinuities, corners, cusps, etc.), where hyperbolicity can deteriorate and shadowing arguments may fail. The presence of singularities actually makes a detailed quantitative analysis of shadowing highly nontrivial. Consider again the cardioid billiard. The cardioid is an example of a non-uniformly hyperbolic system with a cusp singularity. For such a system, shadowing of pseudotrajectories is assured except in small neighborhoods around each point of the singularity set of the Poincar\'{e} map \cite{Serge}. In the case of the cardioid billiard, the singularity set $\mathcal{S}=\mathcal{C}\cup\mathcal{F}$, where $\mathcal{C}$ is the set of initial conditions that start at or will straightaway hit the cusp (i.e., $\mathcal{C}=\{\xi\in\Sigma: s=\pm4\}\cup\{\xi\in\Sigma: p=s/4\}$), and $\mathcal{F}=\{\xi\in\Sigma: p=\pm1\}$ is the set of tangential collision points (``fixed points'') mentioned in Sec.~\ref{modexample}. Roughly speaking, as long as a typical pseudotrajectory does not come too close to any point of the singularity set, the existence of a shadowing trajectory is assured, and the statistical properties of the pseudotrajectory will be very similar to those of the (exact) shadowing trajectory. If and when a pseudotrajectory enters a sufficiently small neighborhood of any point of $\mathcal{S}$ (even only once), shadowing of the whole pseudotrajectory is no longer assured. The main issue then, in the case of the cardioid, is to determine how close to the cusp the particle can come before shadowing breaks down. This is a difficult mathematical problem requiring sophisticated analysis. 
In the absence of more rigorous analysis, it is not possible to make any quantitative statements about how closely typical pseudotrajectories are shadowed by true trajectories and for how long. 

Similar comments apply to the two other billiard systems that were studied. In the case of the stadium billiard, questions concerning shadowability arise when pseudotrajectories pass through sufficiently small neighborhoods of the critical points of the stadium (where the boundary curvature changes discontinuously). In the case of the mushroom, the restriction of the billiard flow to the chaotic region is hyperbolic \cite{BuniMushroom} but shadowing may nevertheless break down near any corner of the mushroom. As before, determining how close to a corner a trajectory must pass before shadowing begins to break down is a difficult mathematical problem.   

To the author's knowledge, shadowing dynamics in the H\'{e}non-Heiles system (which is non-hyperbolic) have not been studied. It has been suggested on the basis of detailed numerical studies (see Ref.~\cite{Henry94}) that any asymptotically-evolved pseudotrajectory of the H\'{e}non-Heiles system at the critical energy does not represent an accurate dynamical history of any single trajectory on the $E=1/6$ energy shell but rather represents a sequence of shorter accurate histories of many different trajectories on the same shell. If (in the absence of any knowledge about shadowing) one accepts this interpretation, then no concrete conclusion concerning the $k$th-nearest-neighbor distance characteristics of the \emph{exact} orbits of the Poincar\'{e} map can be inferred from the numerical experiments of Sec.~\ref{HHexample}. The validity of this interpretation, however, has not yet been firmly established. Thus, the question of whether typical chaotic trajectories on the $E=1/6$ energy shell have distance characteristics consistent with those reported in Sec.~\ref{HHexample}, remains open.

Barring these considerations specific to the systems that were studied, the more fundamental problem is to prove or disprove the following proposition: \emph{For any ergodic component of a 2D non-integrable Hamiltonian flow, typical chaotic trajectories of the Poincar\'{e} map have the same (Euclidean) $k$th-nearest-neighbor distance characteristics as $\mathbf{P}_2$.} It is important to emphasize that the interpoint distances being referred to here are measured using the usual Euclidean metric. 

\subsection{Quantum Chaos}

Strongly-chaotic bounded conservative systems typically possess quantum energy-level nearest-neighbor spacing distributions that are well modeled by the Wigner distribution (see Refs.~\cite{Haake,gutz,Bohigas91,Guhr,Stockmann,Reichl} for examples and discussions). A ``Wigner-like'' energy-level nearest-neighbor spacing distribution (NNSD) has long been widely regarded to be a ``generic'' property of time-reversal-invariant quantum systems having strongly-chaotic classical limits \cite{nogeneric}. The underlying reasons for it being so nevertheless remain elusive. It is \emph{presumed} that, in ``generic'' cases, the Wignerian shape of the NNSD (a property of the quantum eigenvalues) derives solely from the chaoticity of the classical dynamics (a property of the classical trajectories). It is however not fundamentally understood \emph{how} classical chaos is itself responsible for producing the observed Wignerian shape of the energy-level spacing distribution (a point that is often glossed over in the ``quantum chaos'' literature). 

As observed in analyses of the cardioid and stadium billiards (which are strongly-chaotic bounded conservative systems), typical pseudotrajectories of the Poincar\'{e} map have a Wignerian nearest-neighbor distance distribution (NNDD). One of the premises of this paper is that the preceding is a ``generic''  property of strongly-chaotic 2D conservative systems (i.e., a spatial statistical property of the chaotic pseudotrajectories that one expects to observe more generally in numerical simulations of such systems) \cite{reallyWigner}. That the Wigner distribution should have any fundamental significance whatsoever in the description of \emph{classical} chaos is interesting given its significance (albeit as an approximation) in the description of energy-level fluctuations in \emph{quantum} chaos. 

The spectrum of the quantized cardioid billiard is generic in the sense that any sufficiently large sample histogram of the nearest-neighbor spacings is well modeled by the Wigner distribution (see, for example, Ref.~\cite{card}), in accordance with the GOE hypothesis for chaotic systems \cite{BGS}. Incidentally, the higher-order spacings of the cardioid spectrum exhibit clear and significant deviations from GOE predictions (which is expected), but these higher-order spacings likewise do not follow the higher-order ``spacing'' distributions predicted for $\mathbf{P}_2$ (as given by Eq.~(\ref{KNNDDRMT}) with $k\ge2$). So, the correspondence that exists between the NNSD of the quantum energy levels and the NNDD of the classical pseudotrajectories of the Poincar\'{e} map does not extend to the higher-order spacings. 

\section{Conclusion}\label{conc}

A fundamental chaotic process is the deterministic point process (DPP) generated from the successive intersections of a chaotic pseudotrajectory of the flow with a canonical SOS (i.e., a $\Sigma$-type DPP). In the specific context of 2D non-integrable Hamiltonian flows possessing ergodic components, the fundamental question raised in this paper is the following: What are the \emph{spatial statistical} (or \emph{geometrical-statistical}) characteristics of a typical $\Sigma$-type DPP in an ergodic region of phase space? The hypothesis that the (Euclidean) distance characteristics of such a DPP are consistent with those theoretically predicted for the two-dimensional homogeneous Poisson point process ($\mathbf{P}_2$), was put forward and tested. Employing the cardioid and semi-circular mushroom billiards as generic test cases, it was shown that typical chaotic pseudotrajectories of the Poincar\'{e} map have $k$th-nearest-neighbor distance distributions that are well modeled by the corresponding distributions theoretically predicted for $\mathbf{P}_2$. In this respect, the (deterministic) discrete \emph{pseudo}-dynamics generated by the Poincar\'{e} map behave like a Poisson point process in $W$ (the corresponding ergodic component of the SOS), and typical chaotic pseudotrajectories of the map can be viewed as realizations of a Poisson point process in $W$ \cite{WisAll}. An open problem is to determine whether or not the \emph{exact} trajectories of these and other chaotic billiard maps possess spatial statistical properties consistent with those of the \emph{pseudo}trajectories. In principle, a rigorous analysis of the shadowing dynamics is one way of attacking this problem but the presence of singularities introduces highly nontrivial technicalities. 

The interpoint distance characteristics of a $\Sigma$-type DPP initiated and evolved in a non-ergodic region $U$ will, in general, deviate from those of a Poisson point process in $U$. If a $\Sigma$-type DPP is initiated in the chaotic sea, then in certain special cases (e.g., in parameter regimes where the dynamics is strongly, but not fully, chaotic), the deviation will be inconsiderably small, as demonstrated in the case of the H\'{e}non-Heiles system at the critical energy. In general, however, deviations from Poissonian behavior will be significant and a drastically different model is needed to understand the distance characteristics of the chaotic pseudotrajectories. Regular pseudotrajectories will furthermore require the introduction of a different type of stochastic geometric model to understand their corresponding properties. The distance characteristics of $\Sigma$-type DPPs in 2D generic systems is a subject that shall hopefully be explored more extensively in a future publication. 

\begin{acknowledgments} 
The author thanks Eric J. Heller and Tobias Kramer for helpful discussions, and John M. Nieminen for providing the histogram plotting software used to generate some of the figures. The author also benefitted from e-mail correspondence with Nikolai Chernov and Serge Troubetzkoy.\end{acknowledgments}


\begin{thebibliography}{99}
\bibitem{RemPoinMap} Recall that a subset of the continuous trajectories generated by a Hamiltonian flow will intersect a chosen SOS and that successive (oriented) intersections define a discrete dynamics on the chosen SOS that is specified by a Poincar\'{e} return map.
\bibitem{standardMandV} Without loss of generality, the particle is assumed to have unit mass and velocity (i.e., $m=|\mathbf{v}|=1$).
\bibitem{singaset} The singularity set (i.e., the set of initial conditions which exactly start at or immediately hit any corner plus the set of tangential collision points) was excluded when choosing initial conditions.
\bibitem{numdet1} Two error control properties of the ODE solver that required further specification were the absolute and relative tolerances \texttt{AbsTol} and \texttt{RelTol}, respectively. In cases such as the H\'{e}non-Heiles system where the bounded solutions remain small [$\mathcal{O}(1)$], the latter property controls (roughly speaking) the number of correct digits in all solution components. To generate pseudotrajectories of sufficient length [$\mathcal{O}(10^4)$ SOS crossings], the tolerance values \texttt{AbsTol}~$=10^{-10}$ and \texttt{RelTol}~$=10^{-8}$ proved optimal.
\bibitem{nogeneric} There are cases for which the NNSD is qualitatively different from the Wigner distribution (for classic examples, see Refs.~\cite{Delande}). Such cases are deemed to be ``special'' or ``non-generic''.
\bibitem{WisAll} For fully chaotic systems such as the cardioid, $W$ is the entire Poincar\'{e} surface.
\end{thebibliography}
\end{document}